\documentstyle[colordvi,epsfig,12pt,dina4p]{article}
\def\@oddhead{}\def\@evenhead{}
\def\@oddfoot{}
\def\@evenfoot{\@oddfoot}
\def\gev{\rm GeV}

\def\etal{\hbox{$\it et~al.$}}
\def\clb#1 {(#1 Coll.),}
\title{Inclusive $D^0$ and $D^*\pm$ Production in Deep Inelastic $ep$ 
Collisions at HERA}
\author{H1 Collaboration}
\begin{document}

\begin{titlepage}
\begin{flushleft}
%
%
{\tt DESY 96-138} \\
\vspace{1cm}
\end{flushleft}
\vspace*{4.cm}
\begin{center}
\begin{Large}
\boldmath
\bf{Inclusive $D^0$ and $D^{*\pm}$ Production in Deep Inelastic $ep$ 
Scattering at HERA}
\unboldmath

\vspace*{2.cm}
H1 Collaboration 
\end{Large}
\end{center}

\vspace*{1cm}
\begin{abstract}
First results on inclusive 
$\stackrel{\scriptscriptstyle(-)}{\textstyle \;D^{\;0}}$ and $D^{*\pm}$ production 
in deep inelastic $ep$ scattering are reported 
using data collected by the H1 experiment at HERA in 1994.
Differential cross sections are presented for both channels and 
are found to agree well with QCD predictions
based on the boson gluon fusion process. 
A charm production cross section for 
10~GeV$^2\le Q^2\le100$~GeV$^2$ and $0.01\le y\le0.7$ of
$\sigma\left(ep\rightarrow c\overline cX\right)=
(17.4\pm1.6\pm1.7\pm1.4)$~nb is derived.
A first measurement of the charm contribution
$F_2^{c\overline c}\left(x,Q^2\right)$ to the
proton structure function for Bjorken $x$ between $8\cdot10^{-4}$ and
$8\cdot10^{-3}$ is presented. In this kinematic range
a ratio $F_2^{c\overline c}/F_2=0.237\pm0.021{+0.043\atop-0.039}$ is observed.

\end{abstract}
\end{titlepage}
\begin{flushleft}
 C.~Adloff$^{35}$,                  
 S.~Aid$^{13}$,                   
 M.~Anderson$^{23}$,              
 V.~Andreev$^{26}$,               
 B.~Andrieu$^{29}$,               
 R.-D.~Appuhn$^{11}$,             
 C.~Arndt$^{11}$,                 
 A.~Babaev$^{25}$,                
 J.~B\"ahr$^{37}$,                
 J.~B\'an$^{18}$
% ...

\bigskip\bigskip
{\it
% ...
$ ^2$ III. Physikalisches Institut der RWTH, Aachen, Germany$^ a$ \\
% ...
$ ^{11}$ DESY, Hamburg, Germany$^a$ \\
% ...
$ ^{13}$ II. Institut f\"ur Experimentalphysik, Universit\"at Hamburg,
                                     Hamburg, Germany$^ a$  \\
% ...
$ ^{18}$ Institute of Experimental Physics, Slovak Academy of
                Sciences, Ko\v{s}ice, Slovak Republic$^ f$\\
% ...
$ ^{23}$ Physics Department, University of Manchester,
                                          Manchester, UK$^ b$\\
% ...
$ ^{25}$ Institute for Theoretical and Experimental Physics,
                                                 Moscow, Russia \\
 $ ^{26}$ Lebedev Physical Institute, Moscow, Russia$^ f$ \\
% ...
$ ^{29}$ LPNHE, Ecole Polytechnique, IN2P3-CNRS,
                             Palaiseau, France \\
% ...
$ ^{35}$ Fachbereich Physik, Bergische Universit\"at Gesamthochschule
               Wuppertal, Wuppertal, Germany$^ a$ \\
% ...
$ ^{37}$ DESY, Institut f\"ur Hochenergiephysik,
                              Zeuthen, Germany$^ a$\\
% ...
$ ^a$ Supported by the Bundesministerium f\"ur Bildung, Wissenschaft,
        Forschung und Technologie, FRG,
        under contract numbers 6AC17P, 6AC47P, 6DO57I, 6HH17P, 6HH27I,
        6HD17I, 6HD27I, 6KI17P, 6MP17I, and 6WT87P \\
 $ ^b$ Supported by the UK Particle Physics and Astronomy Research
       Council, and formerly by the UK Science and Engineering Research
       Council \\
% ...
$ ^f$ Supported by the Deutsche Forschungsgemeinschaft\\
% ...
   } 
\end{flushleft}
\vspace{1cm}
\centerline{This paper is dedicated to {\bf Hans Duhm} who died im May 1996.}
\newpage


\section{Introduction}

\begin{figure}[h]
  \centering
  \begin{tabular}{ccc}
    \large
    a)&\large b)&\large c)\cr
    \normalsize
    \includegraphics[ bb= 50 50 550 500,width=4.5cm]{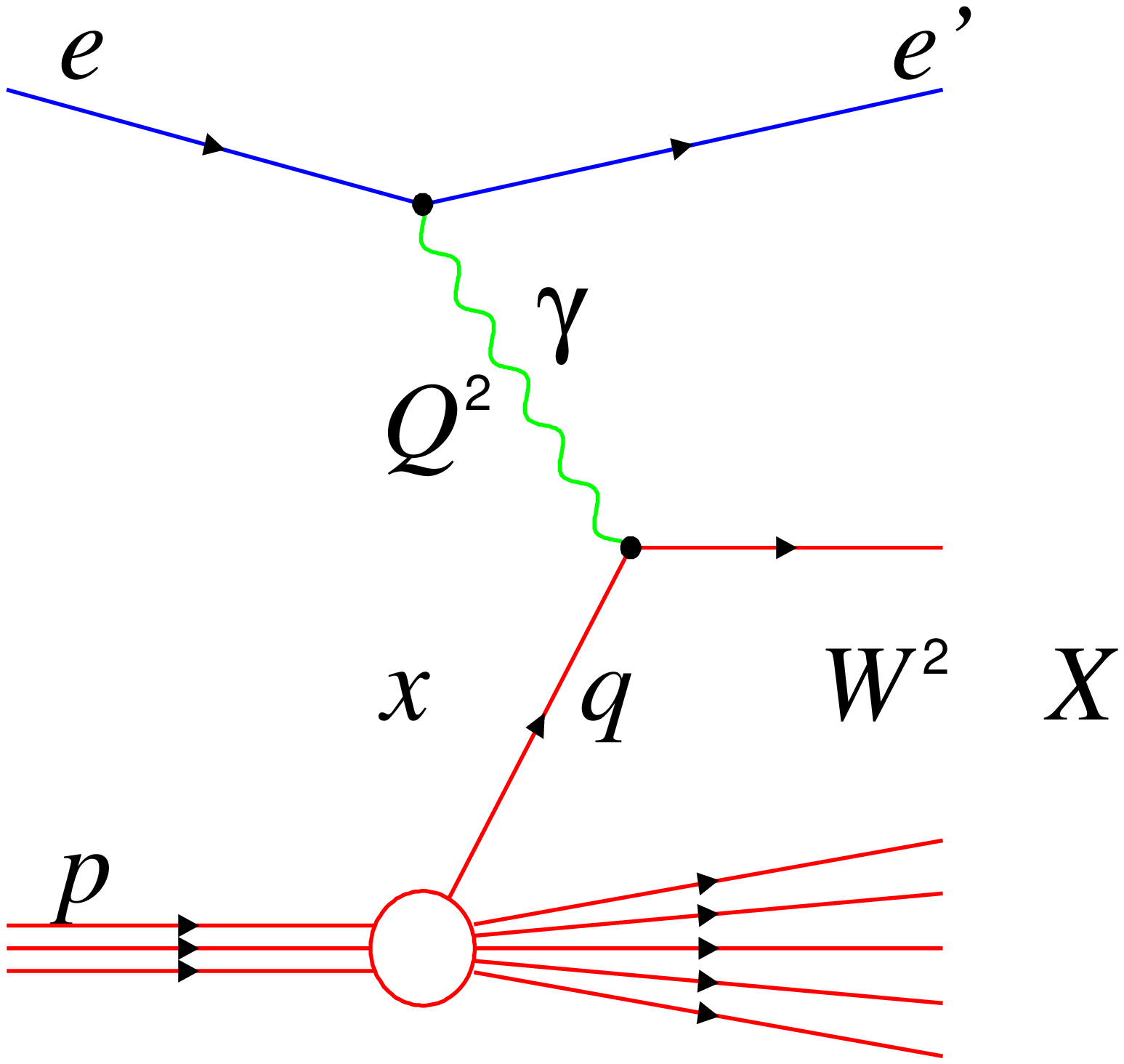}&
    \includegraphics[ bb= 50 50 550 500,width=4.5cm]{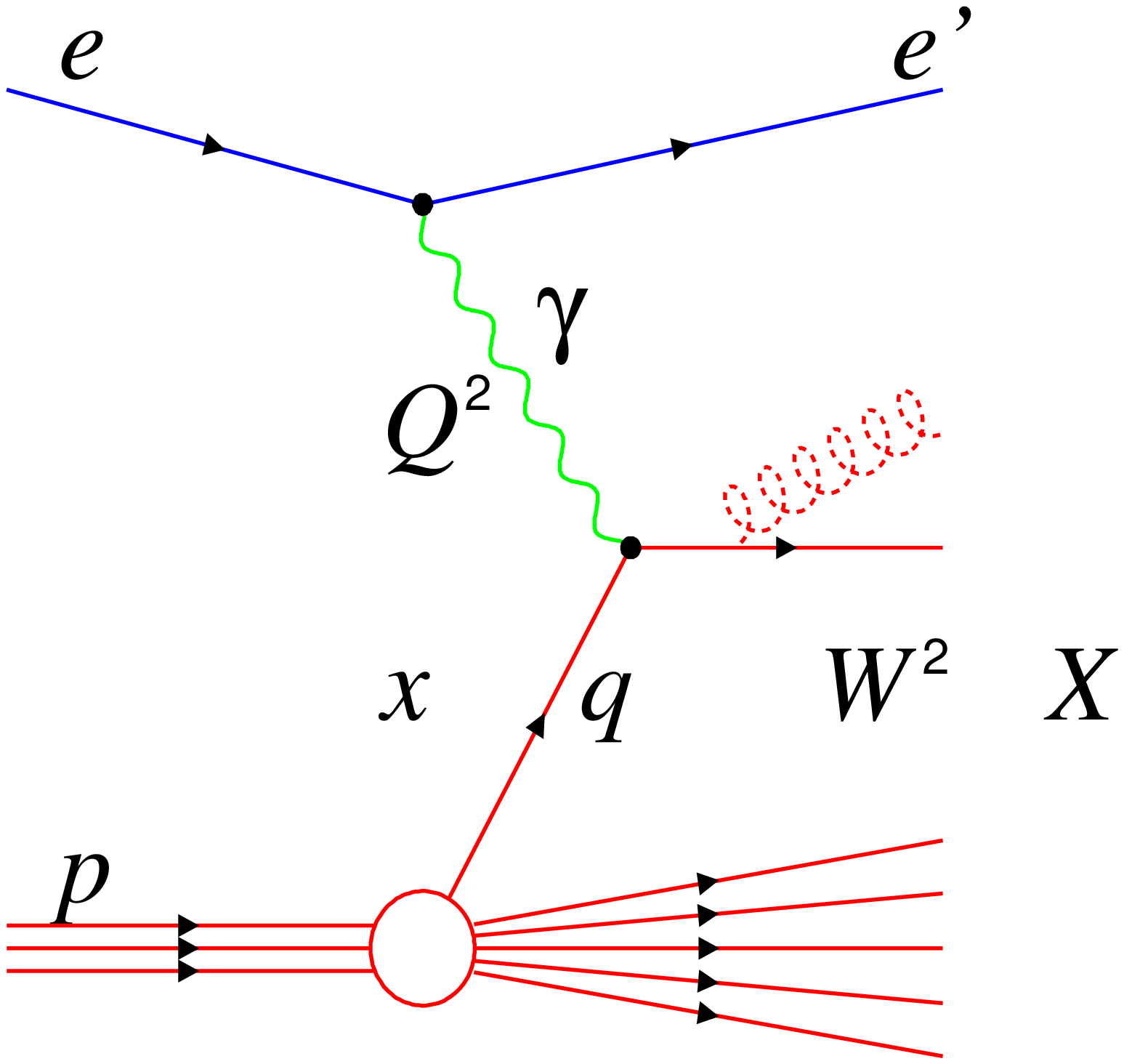}&
    \includegraphics[ bb= 50 50 550 500,width=4.5cm]{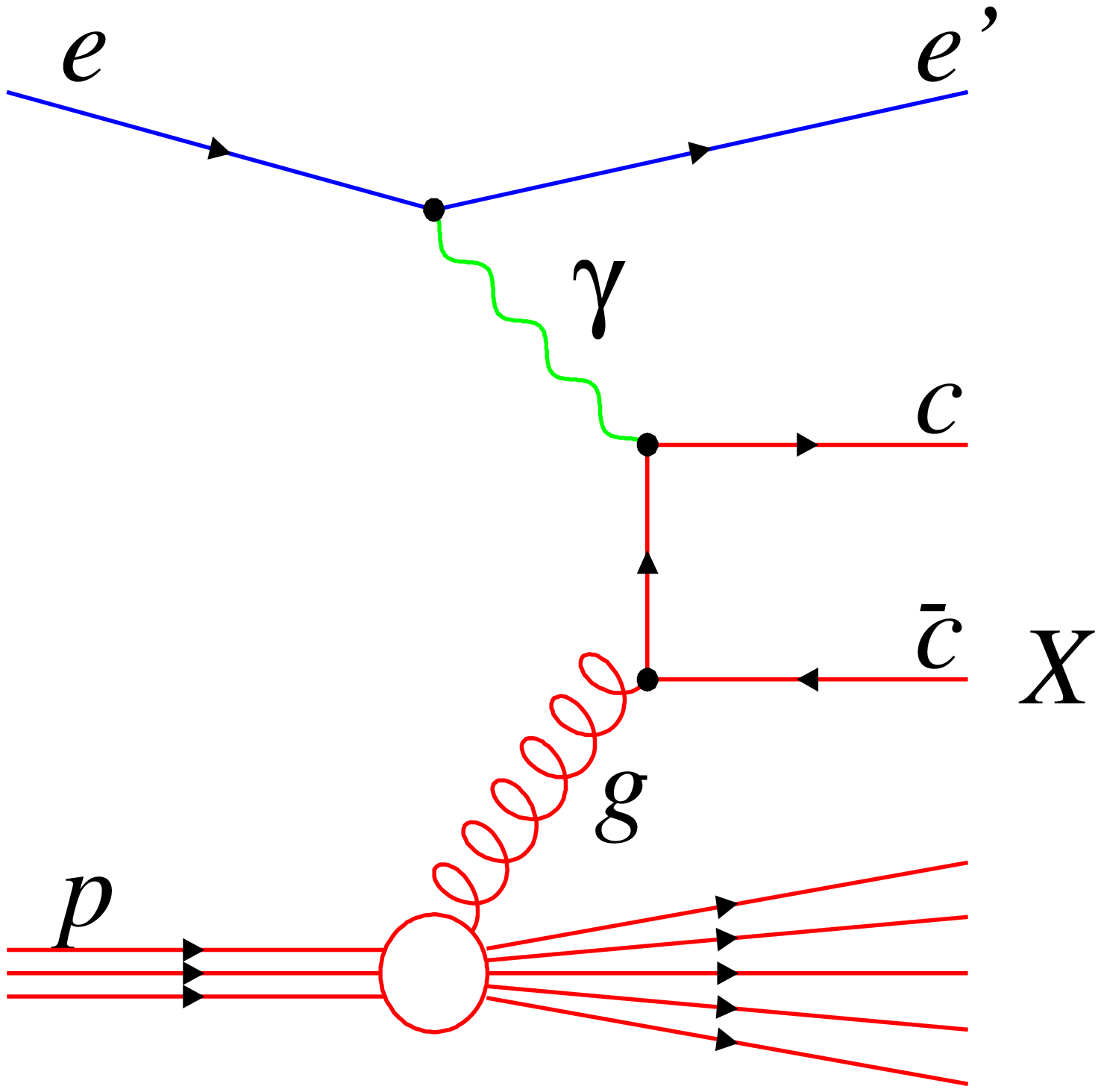}
\cr
  \end{tabular}
\caption[Electroproduction of heavy quarks]%
        {Generic Feynman diagrams for deep inelastic $ep$ scattering 
          up to order
          $\alpha\alpha_s$: (a) $ep$ interaction of the virtual probe 
          ($\gamma,\;Z^0$) with a valence or a sea quark in the proton
          according to the quark parton model (QPM), 
          (b) corrections to process (a) due to gluon radiation off the struck
          quark before or after interacting with the probe 
          (QCD Compton scattering, QCDC), and
          (c) contribution due to boson gluon fusion (BGF).}
\label{DIS}
\end{figure}

The measurement of the charm production cross section in deep inelastic
lepton nucleon scattering (DIS) is of importance for the understanding 
of the parton densities in the nucleon \cite{theory}. 
To order $\alpha\alpha_s$ deep inelastic electron nucleon scattering 
proceeds via the generic diagrams depicted in Fig. \ref{DIS}. 
There is some evidence from the EMC collaboration \cite{emc} that 
heavy quark production should be dominated 
by the {\em boson gluon fusion} process (Fig.\,\ref{DIS}c)
\begin{equation}
 \gamma(Z^0) g \rightarrow c\bar{c},\label{gagcc}
\end{equation} 
which has been calculated in next to leading order \cite{riemersma}.
Charm sea quark contributions (Fig.\,\ref{DIS}a,b), where only a single 
(anti)charm quark is recoiling against the proton,
should contribute only to a small extent. 
Charm production due to fragmentation is suppressed by heavy quark mass 
effects and is therefore expected to be 
small \cite{frag}. For the  $e^+e^-$ annihilation process this expectation 
has been 
confirmed by recent measurements on the $Z^0$ resonance \cite{opal,opalgg}.
Charmed hadron production via the decay
of $b$ flavored hadrons is negligible at HERA, 
because of the small $b$ 
production cross section.
The intrinsic charm model \cite{schuler},  
where 
the proton wave function fluctuates to a $|uudc\overline c\rangle$ state
at the level of a few permille, will give rise to charm production
at large Bjorken $x$. 
Currently the detection of this process is beyond the 
scope of HERA experiments.

A measurement of the charm contribution $F_2^{c\overline c}(x,Q^2)$ to the 
structure function of the proton is supposed to provide 
information on the nature of the charm production process.
Present parton density calculations may include heavy flavors in the proton
via two different routes.
One is to  include charm in the massless quark 
evolution \cite{altarelli}
starting with a charm density of zero at a scale $Q_0^2$ 
of the momentum transfer squared of the virtual photon,
which depends on the mass of the charm quark. This effective procedure is 
presently used in ref.  \cite{mrsd0,mrsh,mrsa}. The second produces
charm  exclusively via boson gluon fusion taking 
into account the charm quark mass \cite{grvho}. 
Although both approaches may lead to very similar predictions for 
 $F_2^{c\overline c}(x,Q^2)$,  the resulting charm quark momentum spectra
will differ. Therefore the momentum 
distribution of the charmed hadrons allows a sensitive test of the charm 
production mechanism.
 
If the photon gluon fusion process is the dominant source of charmed 
hadrons in deep inelastic $ep$ scattering, the observation of 
inclusive charmed hadrons production represents a sensitive probe of the
gluon density $xg(x,Q^2)$ in the proton. The 
measurement of 
$F_2^{c\overline c}(x,Q^2)$ 
can then be used to test different parton density parameterizations.   

In this paper a study of inclusive 
$\stackrel{\scriptscriptstyle(-)}{\textstyle \;D^{\;0}}$
 and $D^{*\pm}$ production 
via the decay channels 
\begin{equation}D^0\rightarrow K^-\pi^+\end{equation} and 
\begin{equation}
D^{*+}\rightarrow D^0\pi^+\rightarrow K^-\pi^+\pi^+\end{equation}
 is presented\footnote{
Henceforth, charge conjugate states are always implicitly included.}.
After a brief description of the 
components of the H1 detector relevant to this analysis (sec. 2) 
the kinematics of inclusive $ep$ scattering is introduced (sec. 3). 
The Monte Carlo simulations used to determine the $D$ meson detection 
efficiencies and to correct
the data are briefly described thereafter (sec. 4). The 
selection of deep inelastic charm events and the possible contributions 
of background sources are then discussed (sec. 5). 
The 
measured $D^0$ and $D^{*+}$ integrated and differential cross sections are 
presented in sec. 6 and compared with QCD based model 
predictions for heavy flavor production in deep inelastic $ep$ scattering. 
Charm production at HERA is also compared with data 
from $\nu N$ scattering.
The charm contribution, $F_2^{c\overline c}(x,Q^2)$, 
to the proton structure function is discussed.
Finally, the results are summarized in sec. 7.        

\section{H1 Detector and Data Sample}
In 1994 HERA has been operated with 820~\gev~protons colliding with 27.6~\gev~
electrons and positrons\footnote{Subsequently electrons will denote both, 
electrons and positrons, respectively.}.
The H1 detector \cite{detector}
is a nearly hermetic multi-purpose apparatus built to investigate the
inelastic high-energy interactions
of electrons and protons at HERA.  
Closest to the interaction point are the
central and forward\footnote {The positive $z$ axis of the H1 coordinate system
is defined by the proton beam direction.} tracking systems, which are
surrounded by a liquid argon calorimeter consisting of an
 electromagnetic and a hadronic section.
A super-conducting solenoid surrounding both the tracking system and the 
calorimeter provides a
uniform magnetic field of 1.15 T parallel to the beam line.
The outermost detector component, the instrumented iron,
    allows the measurement of the hadronic energy leaking out of the
    calorimeter, as well as the identification of muon tracks.

The analysis presented in this paper relies essentially
on parts of the central
tracking system and on the backward electromagnetic calorimeter (BEMC) 
which are described briefly in the following.

The central jet chamber (CJC) consists of two concentric drift chambers
covering
a polar angle of $15^\circ\le\Theta\le165^\circ$. 
It is supplemented by two cylindrical drift chambers at radii of 18 and 47 cm
to determine the $z$ coordinate of the tracks. A
cylindrical proportional chamber is attached to each of the $z$
drift chambers for triggering.

In the backward region,
the polar angle acceptance of the multi-wire proportional chamber (BPC)
covers the range $155^o\le\Theta\le174.5^o$. The reconstructed space point
together with the $z$ vertex position of the event defines the polar angle
of the scattered electron
with a resolution of 1 mrad.

The BPC is attached to the BEMC, 
which covers the polar angular region of $155^o<\Theta<176^o$. 
An energy resolution of $\sigma(E)/E\approx 0.1/
\sqrt{E/\mbox{GeV}}\oplus 0.42\mbox{GeV}/E\oplus 0.03$ is obtained.
The energy
calibration is known to an accuracy of 1\% \cite{bemc}.
A scintillator hodoscope (TOF) situated behind the BEMC is
used to veto proton induced background events based on their early time of
arrival compared with particles from the interaction at the
nominal $ep$ vertex.

This analysis is restricted to DIS events which have a scattered electron 
in the backward region.
These events were
triggered by an electromagnetic shower in the BEMC 
with an energy in excess of 4 GeV which was not
vetoed by an out of time signal in the TOF. The trigger efficiency has been
determined from the data using the redundancy of the H1 trigger system.
For energies of the scattered electron candidate 
$E_e^\prime>13$ GeV,
the trigger efficiency is 100\%.

The luminosity is determined from the rate of the Bethe-Heitler reaction
$ep\rightarrow ep\gamma$. 
The analysis presented in this paper is based on data taken during 1994
running periods with electron and positron beams and corresponds to
a total integrated luminosity of 
${\cal L}_{int}=2.97$~pb$^{-1}$, with an overall uncertainty of
1.5\% \cite{h12}.

\section{Kinematics\label{kine}}
At fixed center of mass energy, $\sqrt{s}$, the
kinematics of the inclusive scattering process $ep\rightarrow eX$ is
completely determined by two independent Lorentz invariant variables, which
may be any two of the Bjorken scaling variables $x$ and $y$, the momentum 
squared $Q^2$ of the virtual boson and the invariant mass squared
$W^2$ of the hadronic final state. In this analysis these variables are 
determined from the measurement of the energy $E^\prime_e$ and the polar 
angle $\Theta_e$ 
 of the scattered electron according to the expressions
(where the electron and proton masses are neglected)
\begin{equation}
\begin{array}{ccc}\displaystyle
Q^2=4E_eE^\prime_e\cos^2\left(
\frac{\Theta_e}{2}\right)&\quad\quad&\displaystyle
y=1-\frac{ E^\prime_e}{ E_e}
\sin^2\left(\frac{ \Theta_e}{2}\right)
\cr\cr\displaystyle
x=\frac{ Q^2}{ ys}&\quad\quad&\displaystyle
W^2=Q^2\left(\frac{ 1-x}{ x}\right)\cr
\end{array}
\end{equation}
where $s=4E_eE_p$, and $E_e$ and $E_p$ denote the energies of the incoming
electron and proton, respectively. Here the scattering angle 
$\Theta_e$ is defined with respect to the proton beam direction.

\section {Monte Carlo Simulation}
Monte Carlo simulation programs
 are used in order to correct the data and to estimate
the systematic uncertainties associated with the measurement.
For the determination of the acceptance of the detector and the $D^0$ and 
$D^{*+}$ selection 
efficiencies, heavy flavor (charm and bottom) DIS events 
are generated 
using the AROMA 2.1 \cite{aroma} program. 
This simulates neutral current heavy quark 
production via the boson gluon fusion process, which is implemented in leading 
order QCD including heavy quark mass effects. A charm quark mass of 
$m_c=1.5\;\gev$ is used. Higher order QCD 
radiation includes initial and final state parton showers (PS)
\cite{ps} using the LEPTO 6.1 code \cite{lepto}. 
JETSET 7.4 \cite{jetset} is used 
to perform the hadronization according to the Lund string model with
the symmetric Lund fragmentation function \cite{lundf}. For the 
gluon density in the proton the MRSH parameterization \cite{mrsh} is used. 
This is compatible with measurements of the proton structure function
at HERA \cite {h193}. 
In order to test the sensitivity of the results to the charm quark mass 
heavy flavor DIS events are also generated with $m_c=1.3\;\gev$ 
and $m_c=1.7\;\gev$.

For the study of any charm contribution to the proton sea,
heavy flavor events are also generated using the LEPTO 6.1 program 
\cite{lepto}. In this generator heavy quark mass effects are not included. 
To avoid the divergences in the QCD matrix elements, 
a cut in the smallest invariant mass $m_{ij}$ of any two partons of an event
is introduced. The difference in the cross section, obtained with this cut,
to the total DIS cross section is then attributed to sea quarks in the proton, 
thus also generating QPM type charm events, where a single
charm quark recoils against the proton remnant.

\section{Event Selection}
\subsection{Selection of Deep Inelastic Scattering Events\label{dissel}}
For the selection of deep inelastic $ep$ interactions the electron candidate is
 identified as the particle giving in the BEMC the most energetic cluster, 
which has also to fulfill the following conditions :
(a) its center of gravity has a radial distance of less than 4 cm from a reconstructed BPC point, and
(b) its lateral shower size has to be smaller than 4 cm. 
 In order to calculate the kinematic variables precisely the position of the 
event vertex is needed. It is defined by at least one well measured track 
crossing the beam axis. 
\begin{figure}[t]
  \centering
  \begin{tabular}{cc}
    \normalsize
    \includegraphics[ bb= 15 20 530 530,width=7.4cm]{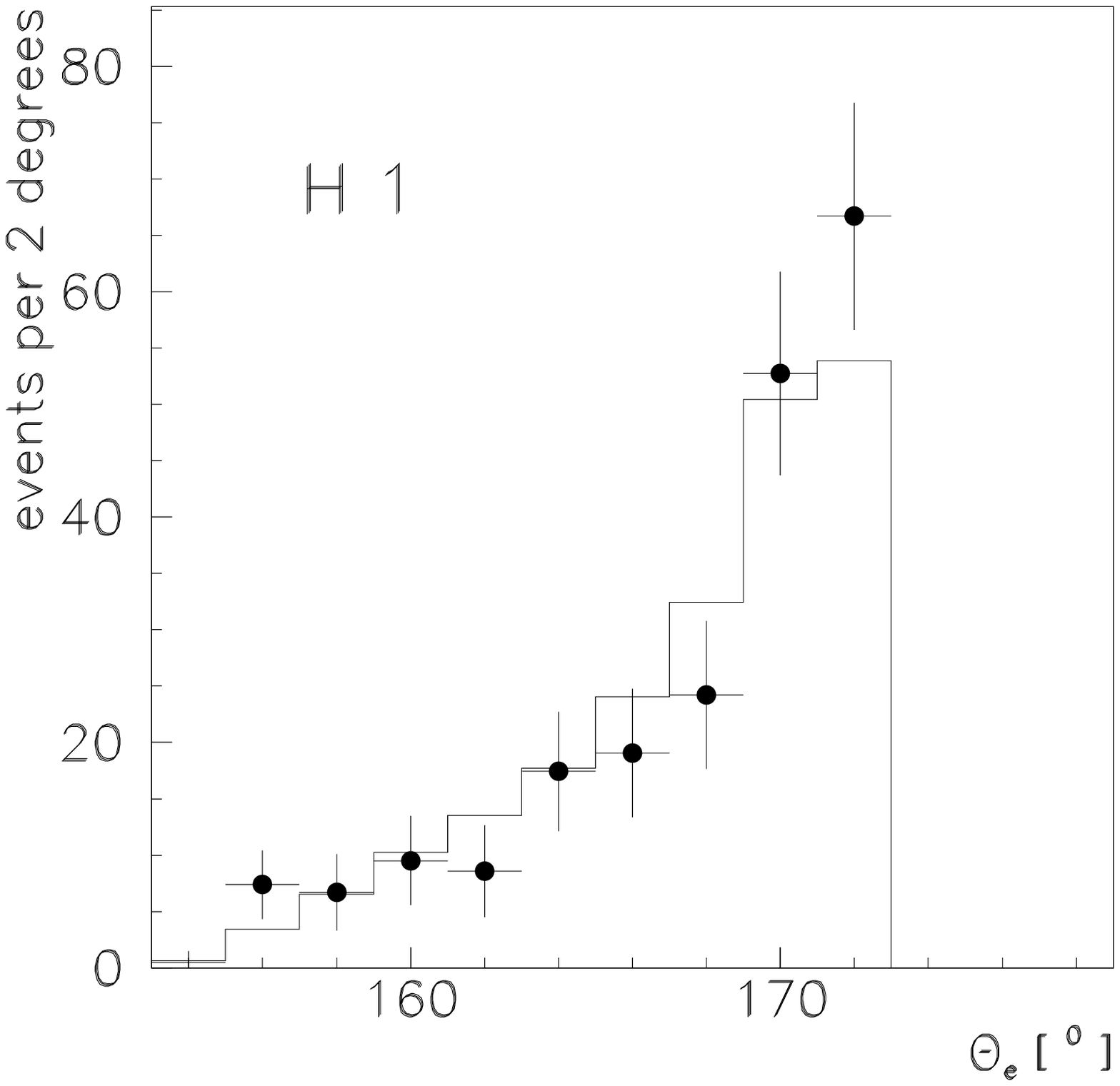}&
    \includegraphics[ bb= 15 20 530 530,width=7.4cm]{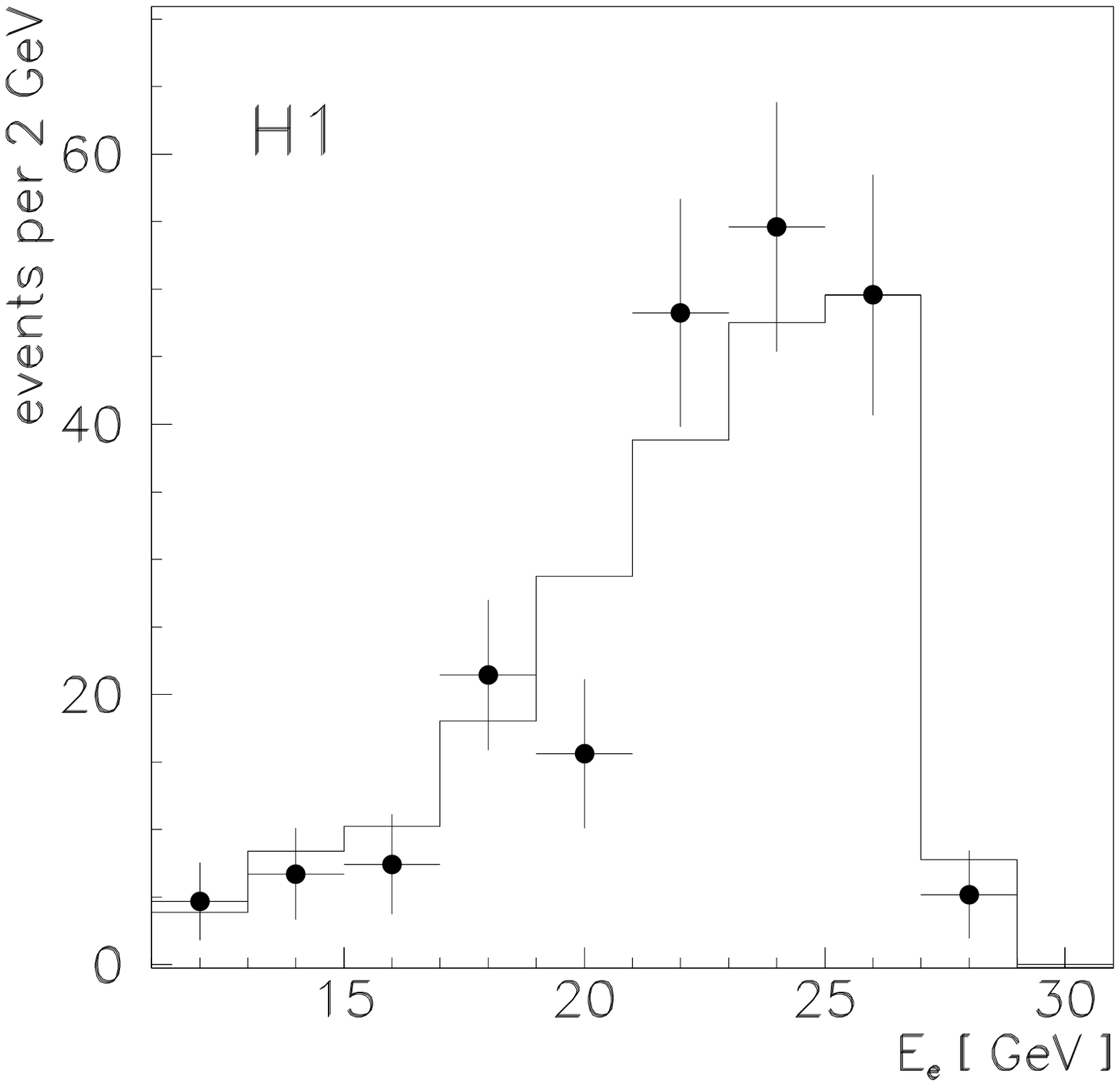}\cr
  \end{tabular}
\unitlength1cm
\begin{picture}(15.,0.1)
\put(1.9,5.3){(a)}
\put(9.3,5.3){(b)}
\end{picture}
\caption{Distribution of (a) the polar angle and (b) the energy of the scattered
electron for charm events as observed in the $D^{*+}$ analysis. The
background subtracted data is compared with the expectation
of the AROMA (histogram) Monte Carlo simulation, normalized to the number of events in the data. 
Only statistical errors are shown.}
\label{scatel}
\end{figure}

The basic kinematic constraints are that  the electron polar angle is in
the range
$155^o\le\Theta_e\le173^o$ and that  the electron energy  
$E_e^\prime>13$ GeV.
The latter requirement ensures that the trigger is fully efficient
 and  
the photoproduction background is small \cite{h12}. 
The analysis is thus restricted to $10~\gev^2<{\it Q^2}<~100~\gev^2$ 
and $y<0.53$. 

The dominant source of non-$ep$ background in the event sample
is due to interactions of the
proton beam with the residual gas and the beam collimators before the
H1 detector.  The level of this background is studied with events originating 
from proton bunches having no colliding electron bunch partner. 
It is found to be less than 0.25\% of the
total number of the selected events.

Events originating from photoproduction, in which the scattered electron escapes
detection but hadronic activity in the BEMC fakes an electron, 
are the only significant $ep$ related 
background. On the basis of Monte Carlo simulations, using the PYTHIA 
generator \cite{jetset}, this background is found to be less than 0.4\%.

Figure \ref{scatel} shows the distribution of the angle and the energy of the
scattered electron for charm events as selected by the $D^{*+}$ analysis in 
comparison to the expectation of the AROMA Monte Carlo simulation. Good 
agreement is observed between the background subtracted data and the 
simulation.

\subsection{Selection of Charm Events}

Reconstructed  $D^0(1864)$ and $D^{*+}(2010)$ mesons are used to tag heavy quark 
production. The analysis is based on charged particles
reconstructed in the CJC. Apart from the electron candidate at least
2 or 3 tracks, depending on the decay mode, have to be fitted to a
common event vertex.
No particle identification is applied so each track may be assumed to be
a pion or a kaon. A minimum track length of 10 cm in the plane 
perpendicular to the beam direction is required for those tracks
considered to form the $D^0$ candidate. 
In both analyses the $K^-\pi^+$ mass combination has to lie 
within the interval $|\eta_{K\pi}|\le1.5$, where $\eta=-\ln\tan\Theta/2$ denotes
the pseudorapidity. 

The $D^0$ is identified via its decay mode
\begin{equation}
D^0\rightarrow K^-\pi^+
\end{equation}
and the $D^{*+}$ through the decay chain
\begin{equation}
D^{*+}\rightarrow D^0\pi^+_{slow}\rightarrow K^-\pi^+\pi^+_{slow}\,.
\label{dstar}
\end{equation}
For the latter use is made of the tight kinematic constraint for the decay of
$D^{*+}\rightarrow D^0\pi^+_{slow}$ \cite{feldmann}. 
A better resolution is expected for the mass difference
\begin{equation}
 \Delta m = m(D^0 \pi^+_{slow}) - m(D^0)\label{deltam}
\label{dstar1}
\end{equation}
than for the $D^{*+}$ mass itself, the
width of which is dominated by the momentum resolution of the detector. 
No cut on the track length is applied in selecting the slow pion.

In addition to the reconstruction of the specific final states the general 
strategy in searching for charm induced events is based on the
hard fragmentation of charmed hadrons \cite{PDG}.
The analysis is performed in the hadronic center of mass ($\gamma^* p$) system.
In analogy to the analyses performed by 
$e^+e^-$ experiments the quantity
\begin{equation}
x_{D}=\left(\frac{|\vec p_{D^0}^{\;*}|}{|\vec p_p^{\;*}|}\right)=
\frac{2\;|\vec p_{D^0}^{\;*}|}{W}\label{xdequa}
\end{equation}
is defined for the $D^0$ in the $\gamma^* p$ frame for both decay
channels\footnote{Quantities defined in the $\gamma^* p$ system are
marked by $^*$.}. 
The mean $\langle x_D\rangle$ of particle combinations originating from
$D^0$ mesons is found to be 
significantly larger than for the combinatorial background.
It therefore may be expected that the 
$D^0$ decay products 
have large momenta in the $\gamma^* p$ system, 
especially in view of the low multiplicity of
the decay mode used in this analysis. 
In order to take advantage of this fact a ranking
of the charged particles is introduced, 
such that the particle of a 
given charge having the largest momentum  in the $\gamma^* p$ system gets the 
rank $R=1$, the 
particle with the second largest momentum gets the rank $R=2$, etc.
Cut are applied in both, the normalized momentum $x_D$ and the rank of the 
particles in the subsequent analyses in order
to suppress the combinatorial background.  

\subsubsection
{Selection of 
$D^{*+}\rightarrow D^0\pi^+_{slow}\rightarrow K^-\pi^+\pi^+_{slow}$
Decays\label{incld*}}
 Only particles having a transverse
momentum of $p_t>0.25\;\gev$ in the laboratory frame are considered for the
reconstruction of $D^0$ candidates. 
The mass combinations $m_{K\pi}$ fulfilling the requirements 
\begin{equation}
\left(R_{K}=1\wedge R_{\pi}\le3\right)
\vee\left(R_\pi=1\wedge R_K\le2\right)
\end{equation}
are calculated for each event. 
The asymmetric cut in the order of the kaon ($R_{K}$) and the pion ($R_{\pi}$)
accounts for the effect of their different masses.
The fractional momentum of the
mass combinations has to satisfy $x_{D}>0.25$. 
At small Bjorken $x$ large $x_{D}$ 
favors $D^0$ candidates going backward in the laboratory frame.
Also those $m_{K\pi}$ mass 
combinations are accepted, which satisfy the condition 
$p_t(K^-\pi^+)>3\;\gev$,
 irrespectively of $R_K,\,R_\pi$ and $x_D$, to retain
the $D^{*+}$ mesons going in the forward direction at small Bjorken $x$.
The $m_{K\pi}$ combinations within $\pm 90$\,MeV of the nominal 
$D^0$ mass of $1.865\,\gev$ are combined with each additional charged 
track with $p_t\ge0.12\,\gev$ and a charge 
opposite to that of the kaon candidate. 
\begin{figure}[t]
\centering
  \begin{tabular}{cc}
    \Large
    \normalsize
    \includegraphics[ bb= 0 140 530 640,width=7cm]{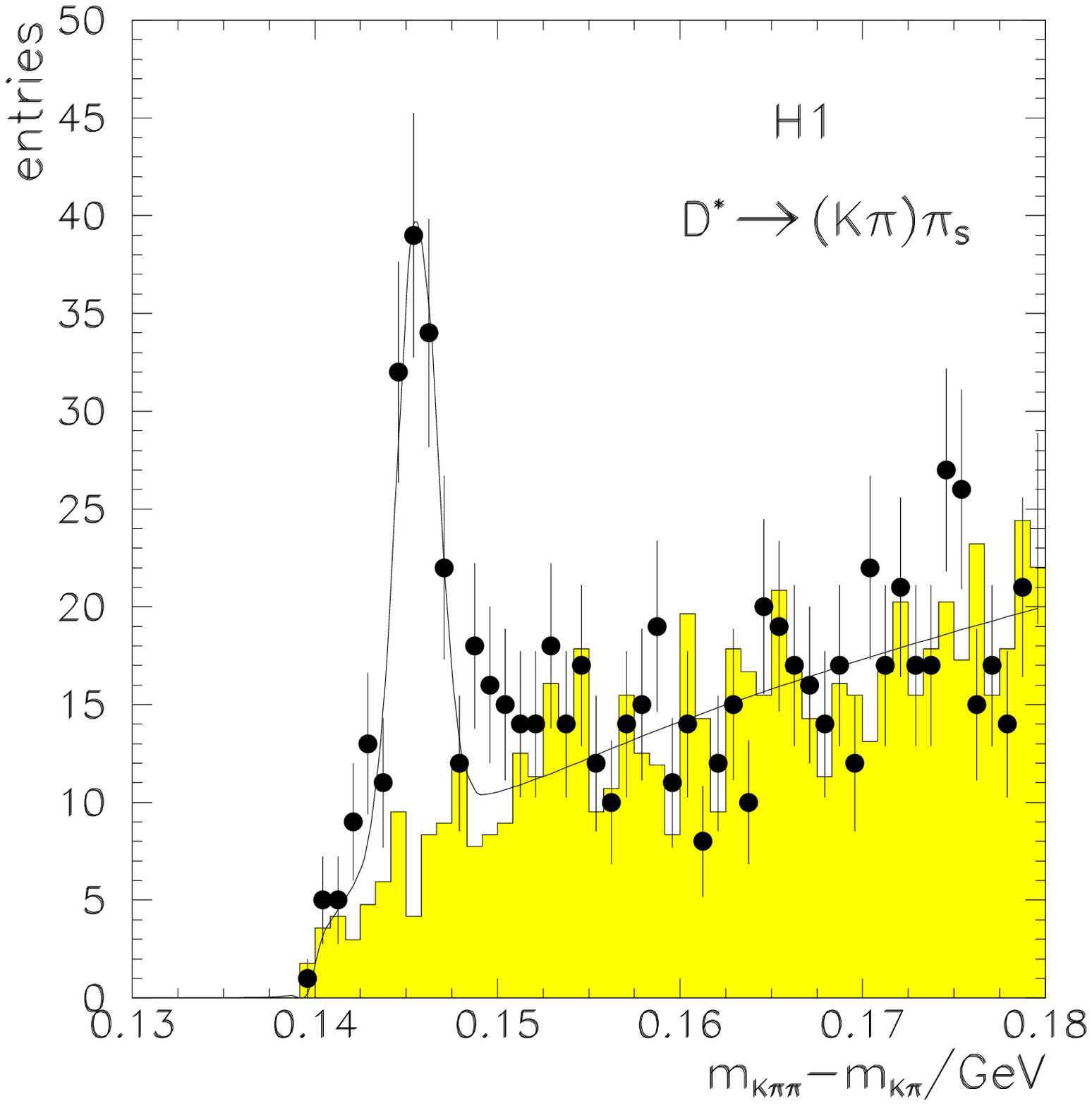}&
    \includegraphics[ bb= 0 140 530 640, width=7cm]{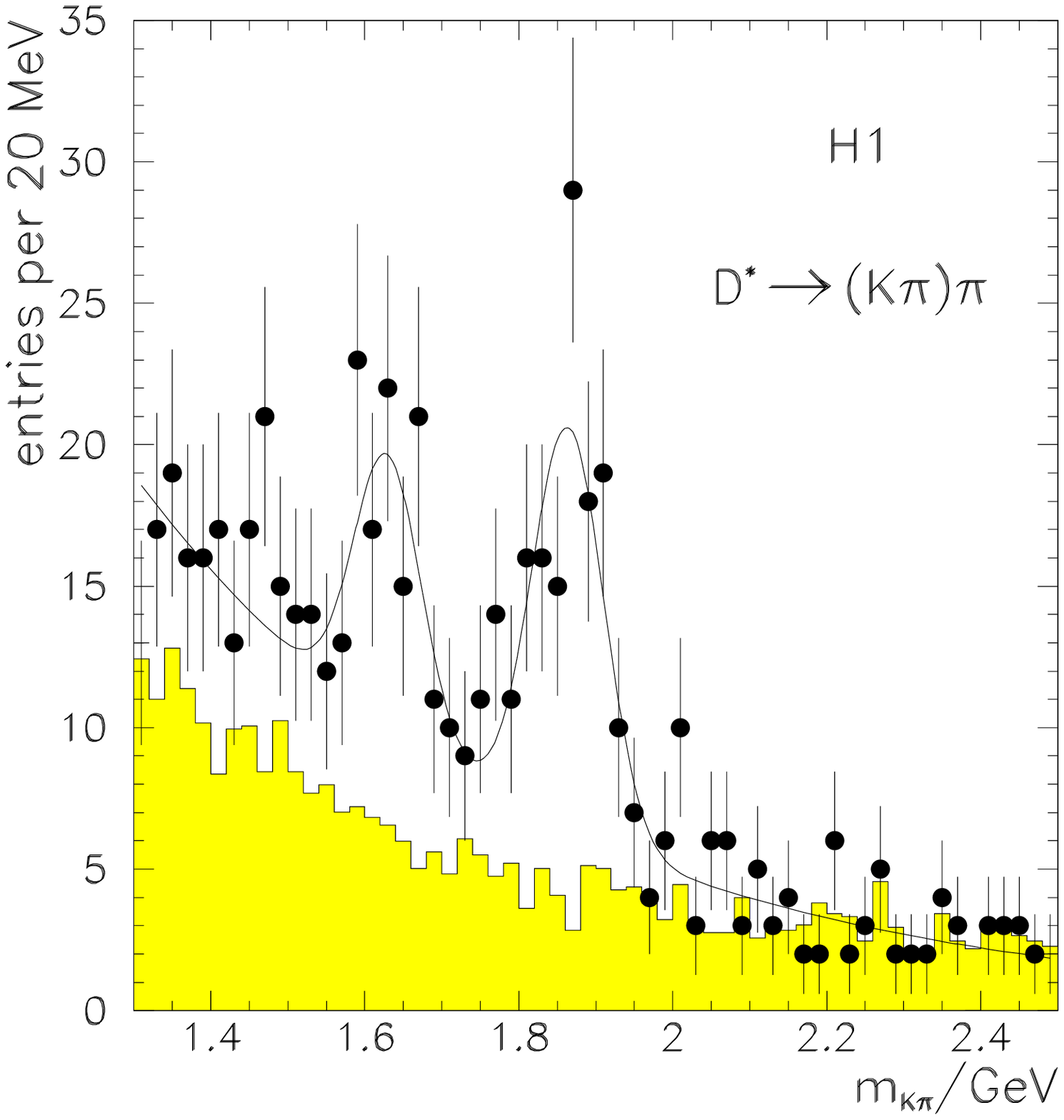}\cr
  \end{tabular}
\unitlength1cm
\begin{picture}(15.,0.1)
\put(2,5.8){(a)}
\put(9.3,5.8){(b)}
\end{picture}

\caption[delta m ]%
    {Distribution of (a) the mass difference 
    $\Delta m$ and (b) the $K^-\pi^+$ mass
    for DIS events selected as described in the text.
    The data points in (a) are obtained from the 
    $K^-\pi^+$ mass combinations fulfilling
    $|m_{K\pi}-m_{D^0}|\le 90$~MeV.
    The shaded histogram shows the background expectation,
    which is obtained from the high mass sideband 
    $2.05\;\gev\le {\it m_{K\pi}}\le2.5\;\gev$ normalized to the region 
    $0.160\;\gev\le\Delta{\it m}\le0.180\;\gev$. 
    The data points in (b) are obtained from the candidates in an
    interval of 2.2 MeV around the nominal $\Delta m$ value expected
    for the decay $D^{*+}\rightarrow D^0\pi^+$. 
    The shaded histogram shows the background 
    expectation from the region 
       $0.170\;\gev\le\Delta{\it m}\le0.180\;\gev$  
       normalized by the two body phase space factors according to the 
       $\Delta m$ intervals. The solid
       lines represent the result of the fits as described in the text.}
\label{massdif}
\end{figure}
 
Figure~\ref{massdif}a shows the distribution of the mass difference for 
accepted $K^-\pi^+\pi^+$ combinations, fulfilling the requirement
$|m_{K\pi}-m_{D^0}|\le 90$~MeV. 
Clear evidence for $D^{*+}$ production is seen in the signal region, 
defined by a $\pm 2.2$\,MeV window around the 
expected mass difference of $m_{D^{*+}}-m_{D^0}=145.4$\,MeV. 
The fit to the data includes
 the $D^{*+}$ signal,  
partially reconstructed $D^0$ mesons 
and the decay modes $D^0\rightarrow\pi^+\pi^-$
and $D^0\rightarrow K^+K^-$, and the two body phase space background.
An inclusive $D^{*+}$ yield of $103\pm13$ is observed.
The peak position at $\Delta m=145.5\pm0.15$ MeV as well as the width of the 
Gaussian of $\sigma=(1.04\pm0.18)$ MeV agree well with the Monte Carlo 
expectations. 
No enhancement is observed in the background sample obtained from mass 
combinations with $2.05\,\gev<{\it m_{K\pi}}<2.5\,\gev$
as shown by the shaded histogram in Fig.\,\ref{massdif}\footnote{
Since the selection is mainly based on the leading particles in the $\gamma^* p$
frame, the background behavior cannot be studied by using 
the like sign $K^-\pi^+$
mass combinations.}.

Figure\,\ref{massdif}b shows the mass distribution of the ${K^-\pi^+}$
combinations for events in the $\Delta m$ signal region.  
The shaded histogram shows the
background expectation from the region 
$0.160\;\gev\le\Delta{\it m}\le0.180\;\gev$
scaled by the two body phase space factors. It describes well the $K^-\pi^+$
mass distribution for $m_{K\pi}>2.05\,\gev$. Below 2.05~\gev~
large differences are observed in the $m_{K\pi}$ distribution 
from the $D^{*+}$
signal region as compared to the background expectation.
These differences are due to contributions where both 
particles originate from the decay $D^{*+}\rightarrow D^0\pi^+_{slow}$.  
Clear signals are observed for the decays $D^0\rightarrow K^-\pi^+$ and 
$D^0\rightarrow K^-\pi^+\pi^0$ where the $\pi^0$ is not detected.
The fit to the data takes into account these two contributions
over an exponentially falling background.
The inclusive $D^0$ yield of $84\pm16$ for the decay $D^0\rightarrow K^-\pi^+$
agrees with the $D^{*+}$ rate obtained from 
the fit to the $\Delta m$ distribution. 
\begin{figure}[t] \centering
    \includegraphics[ bb = 0 10 530 510, width=9cm]{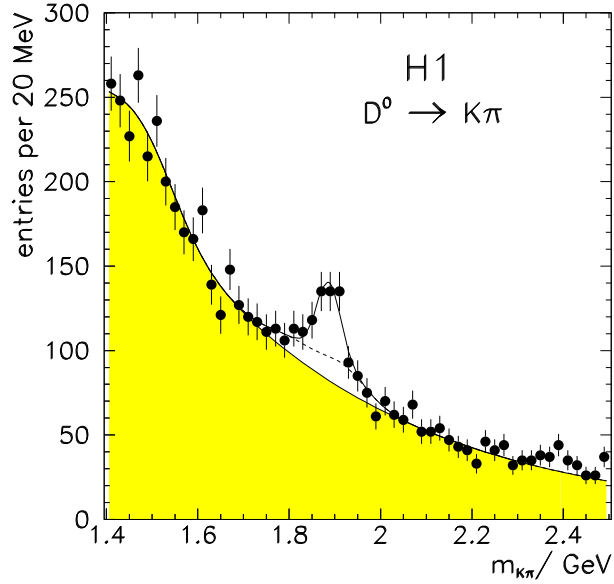}
\caption[D0 Mass distribution]%
        {$K^-\pi^+$ mass distribution observed in the inclusive $D^0$ 
          analysis. The solid line is a fit to the data as indicated
          in the text.
          The dashed line shows the contribution from the wrong $K-\pi$
          assignment to the tracks.
          The shaded area indicates contributions to the fit due to
          combinatorial background and reflections while the 
          dashed line
          describes the mass distribution for the wrong mass assignment
          to the tracks.
         }
\label{mdkp}
\end{figure}

\subsubsection{Selection of $D^0\rightarrow K^-\pi^+$ Decays\label{incld0}}

For each event the ${K^-\pi^+}$ mass combination is calculated only for those 
particles with $p_t\ge1\;\gev$ and fulfilling 
\begin{equation}
R_K=1\wedge R_\pi\le2\;.
\end{equation}
 For the $K^-\pi^+$ mass combination a $p_t>2\;\gev$ and $x_D>0.3$ is required.
Finally a cut in the helicity angle $\Theta^*_K$ of the kaon in
the $D^0$ rest frame with respect to the $D^0$ direction in the laboratory
system of $|\cos\Theta^*_K|<0.5$ is imposed. 
   
Figure \ref{mdkp} shows the mass distribution of the ${K^-\pi^+}$
combinations accepted after these cuts. Clear evidence for $D^0$ production
is observed. The fit to the data contains contributions for the $D^0$ with 
the correct $K,\pi$ assignment to the tracks, for the $D^0$ with the wrong 
$K,\pi$ assignment to the tracks, an exponential distribution 
for the combinatorial 
background, and contributions due to reflections from other $D^0$ decay modes
and from $D^+$ decay at small $m_{K\pi}$.
An inclusive $D^0$ signal of $144\pm19$ is observed within the cuts.
In total $20\pm5$ events are found to be selected by both the $D^0$ and 
$D^{*+}$ analyses. 
 
\section{Results\label{results}}

The results on the differential and the integrated
 cross sections
for the production of charmed hadrons in $ep$ collisions are presented 
for the range 10~\gev$^2\le Q^2\le100~\gev^2$
and $0.01\le y\le0.7$ 
with a small, 7.8\% contribution of 
the extrapolation from the experimentally
accessible range of $y<0.53$ to 0.7.
The cross sections are
calculated from the observed numbers $ N_{obs}$ of $D^0$ and $D^{*+}$
candidates, according to 
\begin{equation}
\sigma(ep \rightarrow D X)=
\frac{N_{obs}}{ {\cal L}_{int} \cdot B \cdot \epsilon_{tot} },
\label{sigman}
\end{equation}
where $\sigma(ep \rightarrow D X)$ is used as an abbreviation for both the
differential and integral cross sections, respectively.
$D$ stands for $D^0$ and $D^{*+}$ and their charge conjugates.
Here ${\cal L}_{int}$ and $B$  refer to 
the integrated luminosity and
the branching ratios $B=B(D^0 \rightarrow K^- \pi^+)=0.0401\pm0.0014$ and 
$B = B(D^{*+} \rightarrow D^0 \pi^+) \cdot B(D^0 \rightarrow K^- \pi^+) =
0.0273 \pm 0.0011$ \cite{PDG}. The quantity $\epsilon_{tot}$ is 
the total efficiency.

The number of observed events, $N_{obs}$, is obtained from the fits to the
spectra.
In the case of differential distributions
the mass distribution for each bin is fit separately by fixing 
the position and the width of
the signals to those values obtained from the total data sample. The effect of
correlations
between the detector resolution and the measured quantity is included in 
the experimental systematic errors.
The AROMA Monte Carlo program 
is used to determine the efficiency arising from 
the selection procedure and the geometrical acceptance of the detector.
A total efficiency, including the acceptance of the apparatus, of 
$(16.3{+1.5\atop-2.1})\%$
and $(5.9{+0.4\atop-0.6})\%$ is obtained for the inclusive $D^{*+}$ and the inclusive 
$D^0$ analyses, respectively.  

The inclusive $D^0$ and $D^{*+}$ cross sections are then converted into a 
charm production cross section in deep inelastic $ep$ scattering 
using the equation
\begin{equation}
\sigma(e p \rightarrow ec {\bar c} X) =
\frac{1}{2}\cdot \frac{\sigma(ep \rightarrow eD X)}
{P(c \rightarrow D)\cdot(1+\xi)}\;,
\label{sigcc}
\end{equation}
where $P(c \rightarrow D)$ denotes the charm quark fragmentation 
probability into a specific $D$ meson and $\xi$ the correction to apply for charm 
production via fragmentation or via $B$ meson decays, respectively.

The fragmentation probabilities $P(c\rightarrow D^0)$ and 
$P(c\rightarrow D^{*+})$ have to be determined from other experiments.
The ARGUS 
\cite{argus} and CLEO \cite{cleo} results from below the $\Upsilon(4S)$ 
together with the HRS \cite{hrs} measurement assuming
$P(b\rightarrow D^{0})=P(c\rightarrow D^{0})$ are
averaged to get 
$P(c\rightarrow D^0)\cdot BR(D^0\rightarrow K^-\pi^+)=0.0205\pm0.0011$.
For $D^{*+}$ mesons the average of all $e^+e^-$ data below the $Z^0$ 
leads to
$P(c\rightarrow D^{*+})\cdot BR(D^{*+}\rightarrow D^0\pi^+)
\cdot BR(D^0\rightarrow K^-\pi^+)=0.0069\pm0.0005$ according to the procedure of 
ref. \cite{opal}\footnote{The value quoted here differs slightly from that 
given in \cite{opal} because it also includes the measurement of ARGUS 
\cite{argus}.}.

The main contribution to the correction $\xi$ arises from gluon splitting. 
Its estimation is based on the OPAL measurement of the mean multiplicity 
per event for $c\overline c$ creation due to gluon splitting with a value
$\langle n_{g\rightarrow c\overline c}\rangle=0.0238\pm0.0048$ 
\cite{opalgg}. This result has been extrapolated to HERA 
energies 
yielding $\xi_{g\rightarrow c\overline c}=0.02\pm0.02$ for $x_D>0.25$.


\begin{table}[t]
\begin{center}
\begin{tabular}{lrr}\hline\noalign{\smallskip}
&$\sigma(ep\rightarrow eDX)$~[nb]
&$\sigma(ep\rightarrow ec\overline c X)$~[nb]
\cr\noalign{\smallskip}\hline\noalign{\smallskip}
$D^0\rightarrow K^-\pi^+$
&$20.4 \pm 2.7{+2.7\atop-2.4}{+1.6\atop-1.2}$
&$19.5 \pm 2.6 {+2.7\atop-2.5}{+1.6\atop-1.2}$
\cr\noalign{\smallskip}\noalign{\smallskip}
$D^{*+}\rightarrow (K^-\pi^+)\pi^+$
&$7.8 \pm 1.0 {+1.2\atop-1.0}\pm0.6$
&$15.1 \pm 1.8 {+2.4\atop-2.0}\pm1.2$
\cr\noalign{\smallskip}\hline\noalign{\smallskip}
Average&&$17.4\pm1.6\pm1.7\pm1.4$
\cr\noalign{\smallskip}\hline\noalign{\smallskip}
\end{tabular}
\end{center}
\caption{Inclusive $D$ meson and charm production cross sections
          in deep inelastic $ep$ scattering for the kinematic range 
          $10\;\gev^2<Q^2<100\;\gev^2$ and $0.01<y<0.7$. The errors refer to
          the statistical error, the experimental systematic, and the model
          dependent uncertainties.}
\label{xsection}
\end{table}

\begin{table}[h]\centering
\begin{tabular}{lccr}\hline\noalign{\smallskip}
Experiment&ref.&Process&$\frac{\sigma(D^{*\pm}X)}{\sigma(D^0X)}$\cr
\noalign{\smallskip}\hline\noalign{\smallskip}
{\bf H1}&&\boldmath$ ep$\unboldmath
&\boldmath $0.38\pm0.07$\unboldmath\cr
&&&\boldmath $\pm0.06$\unboldmath\cr
\noalign{\smallskip}\hline \noalign{\smallskip}
CLEO&\cite{cleo}&$e^+e^-$&$0.48\pm0.06$\cr
ARGUS&\cite{argus}&$e^+e^-$&$0.47\pm0.06$\cr
HRS&\cite{hrs}&$e^+e^-$&$0.47\pm0.06$\cr
DELPHI&\cite{delphi}&$e^+e^-$&$0.43\pm0.05$\cr
ALEPH&\cite{aleph}&$e^+e^-$&$0.36\pm0.04$\cr
OPAL&\cite{opal,opald}&$e^+e^-$&$0.38\pm0.03$\cr
\noalign{\smallskip}\hline \noalign{\smallskip}
NA32&\cite{na32}&$\pi\; Si$&$0.54\pm0.05$\cr
E769&\cite{e769}&$\pi\; Be,Cu,Al,W$&$0.39\pm0.05$\cr
&&$p\; Be,Cu,Al,W$&$0.32\pm0.13$\cr
&&$K\; Be,Cu,Al,W$&$0.24\pm0.08$\cr
\noalign{\smallskip}\hline \noalign{\smallskip}
\end{tabular}
\caption{\label{ratiod*d0}
Experimental Ratio of the inclusive $D^{*\pm}$ cross section over the inclusive $D^0$ cross
section compared to the results from $e^+e^-$ and hadroproduction
experiments. 
}
\end{table}
\subsection{Integrated Cross Sections\label{xsect}}

\begin{table}[h]\centering
\begin{tabular}{lrrr}\hline\noalign{\smallskip}
Model&$m_c\;[\gev]$&Predictions [nb]&This Experiment [nb]\cr
\noalign{\smallskip}\hline\noalign{\smallskip}
GRV   &1.5&11.4&$17.2\pm2.3$\cr\noalign{\smallskip}
MRSH  &1.3&11.3&$17.2\pm2.3$\cr\noalign{\smallskip}
MRSH  &1.5&9.7&$17.4\pm2.3$\cr\noalign{\smallskip}
MRSH  &1.7&8.1&$16.7\pm2.2$\cr\noalign{\smallskip}
MRSA$^\prime$ &1.5&9.7&$17.4\pm2.3$\cr\noalign{\smallskip}
MRSD0$^\prime$ &1.5&8.7&$18.3\pm2.4$\cr\noalign{\smallskip}
\hline\noalign{\smallskip}
H1 $F_2$ fit&1.5&$13.6\pm1.0$&$17.1\pm2.3$\cr\noalign{\smallskip}
\hline\noalign{\smallskip}
\end{tabular}
\caption{Comparison of predicted charm production cross sections with 
         the present measurement for
         $10\;\gev^2\le Q^2\le 100\;\gev^2$ and 
         $0.01<y<0.7$. The theoretical predictions are based 
         on NLO calculations \protect\cite{riemersma} 
         for different parton densities and different 
         values of $m_c$. Also shown is the prediction based on the 
         determination of the gluon density by the H1 NLO fit to the 
         total $F_2$
         data \protect\cite{h12}. The experimental cross section is the 
         average from the $D^{*+}$ and $D^0$ analysis. Total experimental 
         errors are given.
}
\label{model1}
\end{table}

In Tab.~\ref{xsection} the inclusive $D$ meson and charm
production cross sections at the Born level are summarized
for the kinematic range 
$10\;\gev^2<Q^2<100\;\gev^2$ and $0.01<y<0.7$. 
Corrections for QED radiations are applied using
the HECTOR program \cite{hector}. They amount to $5\pm3$~\% for the 
kinematic region explored in this analysis.
The errors in Tab.~\ref{xsection}
reflect the statistical uncertainties, the experimental systematic and the 
model dependent uncertainties due to the extrapolation to the full phase space.
The experimental systematic errors and the model dependent uncertainties
are discussed separately in the subsequent sections.
The results of both measurements are in fair agreement.
Table \ref{xsection} also includes the charm production cross section
obtained by combining the results of the two analyses.

The ratio on the inclusive $D^{*+}$ over the
inclusive $D^0$ cross section is compared with results from $e^+e^-$ 
annihilation and hadroproduction experiments in Tab. \ref{ratiod*d0}.
Agreement is observed with 
the other experiments\footnote{
The ratio from the ARGUS and HRS experiment is calculated using a value
of $BR(D^{*\pm}\rightarrow D^0\pi^\pm)=0.681$ \cite[p. 1171]{PDG} 
instead of 0.55 as quoted in their paper \cite{argus,hrs}. 
}. 
The quoted errors for the present measurement refer to the 
statistical and the experimental systematic error, and take into account 
the systematic uncertainties, common to both analyses. 

In Tab. \ref{model1} the measured charm production cross section 
is compared to the NLO predictions \cite{riemersma} for different 
parameterizations of the gluon density in the proton and different 
values of the charm quark mass $m_c$. 
The results from the inclusive $D^0$ and $D^{*+}$ 
analyses are combined for the determination of the charm production cross 
section. The quoted error refers to the total experimental error, which is
obtained by adding the statistical and the experimental systematic errors
quadratically. 
The measured cross section 
is fairly insensitive to the choice of the model assumptions.
For the theoretical predictions, however, the
integrated charm production cross section shows strong variations
on the model assumptions. 
All theoretical predictions based on the NLO calculations 
\cite{riemersma} are below the measured cross section.  

The table also includes the comparison of the data with the expectation 
of the NLO QCD analysis of the structure function data performed 
by the H1 Collaboration~\cite{h12}. 
In the  evolution  only three light quark flavors are 
taken into account. Heavy quark contributions are dynamically generated 
using the BGF prescription given in ref.~\cite{grv1,grv2}, extended to NLO 
according to ref.~\cite{riemersma}. The scale of the BGF process has been taken as
$\sqrt{Q^2+ 4 m_c^2}$ with a charm quark mass of $m_c=1.5$~\gev. 
The gluon, sea and valence quark distributions were parameterized 
at $Q^2_0=5~\gev^2$. 
The QCD parameter $\Lambda_{QCD}$  is kept fixed 
to the  value of 263 MeV, as determined 
in ref.~\cite{qcdbcd}. 
The parton densities
are derived from a fit of the evolution equations to the 
data \cite{h12,bcdms,nmc}. 
The error on the predicted charm production cross section was obtained 
in propagating the statistical and the uncorrelated experimental systematic
errors on the total $F_2$ data through this fitting 
procedure~\cite{pascaud-lal}.
This determination of the gluon density in the proton yields 
a predicted charm production cross section which comes closer to the result 
of the current analysis.

\begin{table}[t]\centering
\begin{tabular}{lrr}\hline\noalign{\smallskip}
Experimental&&\cr
Systematic Errors&$D^0$&$D^{*+}$\cr
\noalign{\smallskip}\hline\noalign{\smallskip}
reflections,&&\cr
background shape\&&&\cr
mass resolution&$\pm0.10$&$\pm0.11$\cr\noalign{\smallskip}
wrong $K^-\pi^+$ assignment&$\pm0.0$3&-\cr\noalign{\smallskip}
tracker efficiency&${+0.06\atop-0.02}$&${+0.09\atop-0.03}$\cr\noalign{\smallskip}
luminosity&0.015&0.015\cr
$\gamma p$ contribution&$<$0.004&$<$0.004\cr\noalign{\smallskip}
radiative corrections&$\pm0.03$&$\pm0.03$\cr\noalign{\smallskip}
branching ratios&$\pm0.03$&$\pm0.04$\cr
\noalign{\smallskip}\hline 
\noalign{\smallskip}
total $ep\rightarrow eDX$&${+0.13\atop-0.12}$
&${+0.15\atop-0.13}$\cr\noalign{\smallskip}\hline
\hline\noalign{\smallskip}
\end{tabular}
\caption{Summary of the relative experimental systematic uncertainties of the
integrated inclusive cross section.}
\label{table}
\end{table}

\subsubsection{Experimental Systematic Uncertainties 
\label{uncert}}

The experimental systematic uncertainties are summarized in Tab.~\ref{table}. 
By adding the different contributions in quadrature 
total relative errors of ${+13\%\atop-12\%}$ and ${+15\%\atop-13\%}$ 
are found for the inclusive $D^0$ and $D^{*+}$ cross sections, 
respectively.

In the case of the inclusive $D^0$ analysis the dominant experimental
systematic uncertainty arises from
the description of the combinatorial background and the shape 
of the reflections below the $D^0$ mass. A simple phenomenological
ansatz has been used to parameterize these contributions in the fit
shown in Fig. \ref{mdkp}.
The uncertainty introduced by this procedure is estimated by varying the
$m_{K\pi}$ range in which the data is fitted.
The form of the wrong $K-\pi$ assignment and its rate with respect 
to the correct $K-\pi$ assignment is studied by Monte Carlo simulation.
The given error is determined from this simulation.

For the inclusive $D^*$ analysis, the uncertainties coming from the shape of
the combinatorial background are determined from the $\Delta m$ 
distribution of the 
high $m_{K\pi}$ sideband, which matches well the two body
phase space expectation. 
The given uncertainty 
accounts for the fact that the shap of the combinatorial background may
differ from the two body phase space prediction. The contribution due 
to reflections in the $D^0$ mass window is studied with Monte Carlo methods.

For the differential cross sections in the subsequent sections an error of
10\% due to the procedure used to determine the inclusive $D$ meson yields 
per bin is added quadratically to the systematic uncertainties.

The uncertainty of the single track reconstruction efficiency 
are investigated as described in ref. \cite{ccgp}.


\begin{table}[t]\centering
\begin{tabular}{lrr}\hline\noalign{\smallskip}
Model&&\cr
Uncertainties&$D^0$&$D^{*+}$\cr
\noalign{\smallskip}\hline\noalign{\smallskip}
charm fragmentation&&\cr
function&$\pm$0.04&$\pm$0.04\cr\noalign{\smallskip}
GRV-HO Gluon&-0.01&-0.01\cr
MRSD0${^\prime}$ Gluon&$+0.05$&$+0.05$\cr
Gluon from H1 $F_2$ fit&-0.01&-0.02\cr\noalign{\smallskip}\hline
\noalign{\smallskip}
$m_c=1.3\;\gev$&$\pm0.04$&+0.03\cr
$m_c=1.7\;\gev$&$\pm0.04$&-0.06\cr\noalign{\smallskip}
\hline 
\noalign{\smallskip}
total $ep\rightarrow eDX$
&${+0.08\atop-0.06}$&$\pm0.07$\cr\noalign{\smallskip}\hline
\noalign{\smallskip}
fragmentation \&&&\cr
gluon splitting&$\pm$0.02&$\pm$0.02\cr\noalign{\smallskip}
$b$ contribution&$<$0.02&$<$0.02\cr
\hline 
\noalign{\smallskip}
total $ep\rightarrow ec\overline cX$
&${+0.08\atop-0.06}$&$\pm0.08$\cr\noalign{\smallskip}\hline
\hline\noalign{\smallskip}
\end{tabular}
\caption{Summary of the relative model dependent uncertainties of the
 inclusive $D$ meson and charm production cross sections.}
\label{model}
\end{table}

\subsubsection{Model Dependent Uncertainties}
Various sources of model dependent uncertainties of the cross sections 
have been investigated and
are summarized in Tab.~\ref{model}.
They are based on estimated errors 
on the total efficiency due to the uncertainties in the
charm fragmentation function and its QCD evolution, 
in the gluon density of the proton, and in the effect of 
the charm quark mass on the inclusive $D$ cross section. 
In the case of the charm production cross section, they also include 
an error for
the estimate of the correction factor $\xi$. 
The total model dependent error is obtained by adding the different 
contributions quadratically.

The influence of the charm fragmentation function on the cross section
is checked by using different parameterizations \cite{lundf,cfrag}.
A change in the total efficiency of less than 
1\% is found. The effect of the QCD evolution is studied by shifting the mean 
$\langle x_D\rangle$ by 0.1 which corresponds roughly to the difference in 
the gluon radiation by changing $W$ by more than one order of magnitude.
This results in an uncertainty of 4\%.

The sensitivity of the measured cross section on the parton density in the
proton has been studied by using different parameterization. In
addition to the MRSH, MRSA$^\prime$, MRSD0$^\prime$ and GRV-HO 
parameterizations also the gluon density resulting from the NLO QCD fit to 
the total $F_2$ measurement from H1 \cite {h12} has been used 
for efficiency calculations.
Only small variations in the total efficiency are observed.

In order to investigate the influence of the charm quark mass the correction
factors have been calculated by using the AROMA Monte Carlo program with
$m_c=1.3\;\gev,~1.5\;\gev$, and $1.7\;\gev$. Again only 
small variations in the efficiency are observed.

\begin{figure}[t] 
\centering
\includegraphics[ bb= 0 20 530 530, width=9cm]{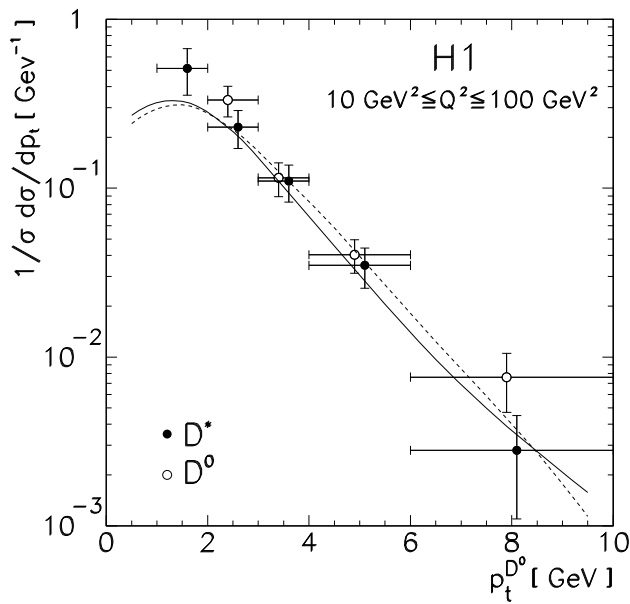}
\caption{Normalized distributions of 
        $1/\sigma\;d\sigma/dp_t(D^0)$ of the $D^0$ mesons from the
        $D^{*+}$ (full points) and the $D^0$ analysis (open circles) in
        comparison to the AROMA expectation for $m_c=1.3\;\gev$
        (full line)
        and $m_c=1.7\;\gev$ (dashed line) using the MRSH 
        parameterization of the parton densities in the proton.
        }
\label{pt}
\end{figure}

\subsection{Differential Cross Section\label{diffsect}}

Figure \ref{pt} shows the normalized differential cross 
section $1/\sigma\; d\sigma/dp_t(D^0)$ for  inclusive $D^0$ and for $D^0$
via $D^{*+}$ 
production for $10\;\gev^2<Q^2<100\;\gev^2$ and $0.01<y<0.7$.
The normalization is done with respect to the integrated 
cross section. Only statistical errors are shown.
No source of systematic 
uncertainties is found which would introduce additional uncorrelated 
systematic errors.
Good agreement between
 the normalized $p_t(D^0)$ spectra of the two analyses is 
observed\footnote{Although the $D^0$ produced via $D^{*+}$ 
decay does not originate 
directly from the charmed quark, no differences between the $p_t(D^0)$ spectra
for the two analyses 
are expected due the smallness of the kinetic energy available in the
$D^{*+}\rightarrow D^0\pi^+$
decay.}. The data are also compared with the expectation from the boson gluon 
fusion process
for charm quark masses of $m_c=1.3\;\gev$ 
and $m_c=1.7\;\gev$. 
The measured differential distributions are
 in good agreement with the expectation.
The differences due to the choice of the charm quark mass
are much smaller than the present accuracy of the data.
The data have also been compared to calculations using the 
GRV-HO \cite{grvho} and MRSD0$^\prime$ \cite{mrsd0} gluon
densities, which lead to 
 normalized $p_t(D^0)$ distributions nearly identical with the one obtained for 
the MRSH parameterization.  

\subsection{Mechanisms of Charm Production in $ep$ Scattering\label{fragment}}

\begin{figure}[t]
\centering
\includegraphics[ bb= 0 0 530 530,width=9cm]{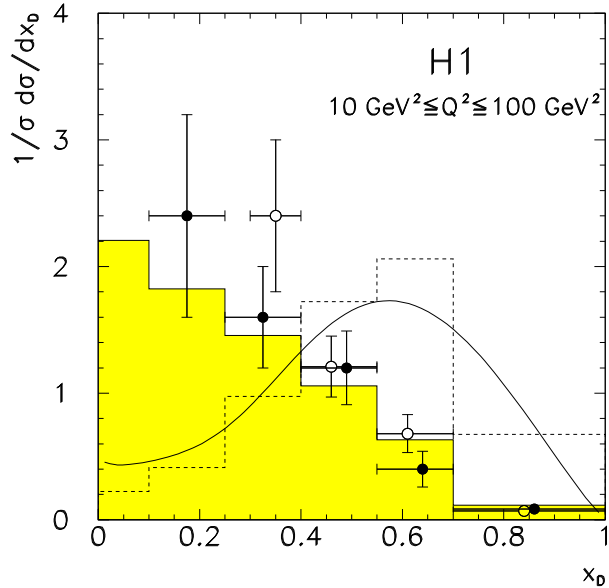}
\caption{ Normalized $x_D$ distribution in deep inelastic 
          $ep$ scattering 
          at $\langle W\rangle\approx 125\;\gev$ for $|\eta_D|<1.5$.
          The open points represent the $D^0$ data, 
          the closed points the $D^*$ data.          
          The shaded histogram shows the expectation of the boson gluon fusion
          process according to the AROMA Monte Carlo
          simulation. The dashed histogram
          shows the expectation from a charm sea contribution, calculated by
          selecting QPM events with the
          LEPTO/MEPS Monte Carlo program. 
          The full line gives the result of the
          QCD evolution of the 
          $\stackrel{\scriptscriptstyle(-)}{\textstyle \nu}N$ data.}
\label{xd}
\end{figure}

The charm production mechanisms in $ep$-scattering can be investigated 
by comparing 
fully corrected data with  expectations 
for charm production off a charm quark in the proton sea 
and via boson
gluon fusion 
(BGF).
It is assumed that the fragmentation function $D^c_D(z)$
with $z=E_D/E_c$, the ratio of the charmed meson energy $E_D$ to the charm quark
energy $E_c$, is a universal function independent of the charm creation
mechanism. This has been shown in ref.~\cite{kk} to hold for both 
electron-positron annihilation
and neutrino-nucleon scattering \cite{cdhs,e531} provided the data is
corrected for QCD radiation \cite{altarelli}. 
Figure \ref{xd} shows
the corrected distributions ${1}/{\sigma}\; {d \sigma}/{dx_D}$, 
which are a convolution of the charm production spectrum with the
fragmentation function in comparison with: 
\begin{enumerate}

\item Monte Carlo simulated data using the AROMA generator in which charmed
quarks are produced by BGF and fragmented according to  JETSET.   

\item Measurements of
$\stackrel{\scriptscriptstyle(-)}{\textstyle \nu}N$-scattering 
by the CDHS \cite{cdhs} and E531 \cite{e531} experiments, where
charm production is expected to proceed mainly
via $W^\pm$-scattering off a strange sea quark in the proton\footnote{In
case of the $\nu N$ scattering, the data also include sizeable contributions 
from valence quarks.} 
(in analogy to Fig. 1a).
It can be shown that the variable $z$, measured in the proton rest frame,
approximately transforms into $x_D$ in the $W^*p$ system.
The results are evolved from $\langle W\rangle=8\;\gev$ to 
$\langle W\rangle=125\;\gev$~\cite{altarelli}.


\item Monte Carlo simulated data using the LEPTO 6.1 generator,
from which only
charm sea quark events are selected. 
The fragmentation is modelled using JETSET. 
\end{enumerate}

The data are restricted to the region $|\eta_D|<1.5$ in order to avoid large 
$x_D$ dependent correction factors due to the detector
acceptance.
The total error is                                               
obtained by adding in quadrature the statistical error for the real and 
simulated data 
and the sum of the model dependent systematic uncertainties.
The measurement errors are dominated by statistical effects. No source of
 systematic uncertainties is found which would introduce
significant uncorrelated errors. Good agreement is observed between the $x_D$
distribution obtained from  the two different analyses.

\noindent The differences between 
the expectation 
from the BGF model, where two charm quarks are recoiling against
the proton remnant in the hadronic center of mass system, on the one hand 
and the predictions from the 
$\stackrel{\scriptscriptstyle(-)}{\textstyle \;\nu}N$
 data and the quark parton model, 
where only one charm quark emerges opposite to 
the proton remnant, on the other hand
are evident. The BGF model agrees very well with the data which justifies 
the usage of this model throughout the paper.

To determine the fraction of the
charm production cross section which may be attributed to a photon
scattering off charm sea quarks in the proton, a superposition of the 
boson gluon fusion and the sea quark predictions, i.e. a function of the shape 
\begin{equation}
1/\sigma d\sigma/dx_D = (1-\epsilon)(1/\sigma d\sigma/dx_D)_{BGF}
+\epsilon (1/\sigma d\sigma/dx_D)_{sea}
\end{equation}
is fitted to the $x_D$ distribution  of Fig. \ref{xd}. 
Depending on the form of a charm sea quark contribution, i.e.
using the results of the LEPTO Monte Carlo generator or 
the $\stackrel{\scriptscriptstyle(-)}{\textstyle \;\nu}N$
data, and depending on the charm quark
mass $m_c$, the fit yields values for  $\epsilon$ between $-0.062\pm0.035$
and $-0.041\pm0.031$. According to the Bayesian approach \cite{PDG}
these values are converted to an upper limit on a charm sea quark contribution
to charm production in deep inelastic $ep$ scattering in the kinematic
range at HERA of 
\begin{equation}
\epsilon<0.05
\end{equation}
at the 95\% confidence level. 
It may thus be assumed that boson gluon fusion is 
the dominant charm production process in DIS at HERA.

\subsection{Charm Contribution to the Proton Structure Function\label{f2c}}

\begin{table}[t]\centering
\begin{tabular}{crrrr}\hline\noalign{\smallskip}
\noalign{\smallskip}
Mode&$\langle Q^2\rangle$&$\langle x\rangle$&$F_2^{c\overline c}$
&$F_2^{c\overline c}/F_2$\cr
&$[\gev]$&&&\cr
\noalign{\smallskip}\hline\noalign{\smallskip}
$D^*$&12&.0008&$0.211\pm0.049\;{+0.045\atop-0.040}$
&$0.198\pm0.045\;{+0.039\atop-0.033}$
\cr\noalign{\smallskip}
$D^0$&12&.0020&$0.263\pm0.036\;{+0.043\atop-0.041}$
&$0.304\pm0.043\;{+0.045\atop-0.042}$
\cr\noalign{\smallskip}
$D^*$&12&.0032&$0.190\pm0.054\;{+0.052\atop-0.049}$
&$0.254\pm0.072\;{+0.063\atop-0.058}$
\cr\noalign{\smallskip}
\noalign{\smallskip}\hline \noalign{\smallskip}
$D^*$&25&.0008&$0.324\pm0.099\;{+0.065\atop-0.058}$
&$0.244\pm0.075\;{+0.044\atop-0.038}$
\cr\noalign{\smallskip}
$D^0$&25&.0020&$0.253\pm0.069\;{+0.041\atop-0.040}$
&$0.248\pm0.086\;{+0.035\atop-0.033}$
\cr\noalign{\smallskip}
$D^*$&25&.0032&$0.222\pm0.066\;{+0.044\atop-0.039}$
&$0.255\pm0.076\;{+0.045\atop-0.038}$
\cr\noalign{\smallskip}
\noalign{\smallskip}\hline \noalign{\smallskip}
$D^*$&45&.0020&$0.156\pm0.070\;{+0.031\atop-0.028}$
&$0.127\pm0.059\;{+0.023\atop-0.020}$
\cr\noalign{\smallskip}
$D^0$&45&.0032&$0.275\pm0.074\;{+0.045\atop-0.042}$
&$0.249\pm0.071\;{+0.035\atop-0.033}$
\cr\noalign{\smallskip}
$D^*$&45&.0080&$0.200\pm0.064\;{+0.040\atop-0.035}$
&$0.269\pm0.088\;{+0.047\atop-0.039}$
\cr\noalign{\smallskip}
\noalign{\smallskip}\hline \noalign{\smallskip}
\end{tabular}
\caption{The value of
         $F_2^{c\overline c}$ and of the ratio $F_2^{c\overline c}/F_2$. 
         The errors refer to the statistical and the experimental systematic
         errors.\label{tf2c}}
\end{table}

The charm contribution $F^{c\overline c}_2(x,Q^2)$ to the structure function
is obtained by using the expression for the  
one photon exchange cross section for charm production
\begin{equation}
\displaystyle
\frac{d^2\sigma^{c\overline c}}{dxdQ^2}=\frac{2\pi\alpha^2}{Q^4x}
\left(1+\left(1-y\right)^2\right)\;F^{c\overline c}_2(x,Q^2)\;,
\end{equation} 
with the simplification that the Callan-Gross relation holds,
i.e. $R=F_2/2xF_1-1=0$. 
According to QCD calculations a maximum 
value of $R\approx0.38$ is expected in the kinematic range explored in the
present analysis. An increase in $F^{c\overline c}_2(x,Q^2)$ 
of at most 2\% is obtained
 at small $x$ for $R=1$. 

The charm contribution to the proton structure function is obtained from the
numbers of reconstructed $D^{*+}$ and $D^0$ mesons, 
which are converted to bin
averaged cross sections according to Eqns. (\ref{sigman},\ref{sigcc}) and using
the Monte Carlo efficiency calculation. The coarse binning in $x$ and $Q^2$ is 
dictated by the small statistics available. 
The bin averaged cross section is 
corrected for QED radiative effects using the program HECTOR 
\cite{hector}. 

In Tab.~\ref{tf2c} the results of the $F_2^{c\overline c}$ measurements are 
summarized. The errors refer to the statistical and to the experimental 
systematic uncertainties. 
The systematic uncertainties are summarized in Tab.~\ref{table3}.
They include an
error due to the angular 
and energy resolution of the electron measurement, which is dominated by
the uncertainty of the BEMC energy calibration. A shift in the energy 
scale of 1\% 
introduces uncertainties of 6\%~(9\%) in $F_2^{c\overline c}$ for the 
low (high) $x$ bin in the $D^{*+}$ analysis at 
$\langle Q^2\rangle=12\;\gev^2$
due to bin migrations in $x$. For all other bins this effect is 
below 5\%. The experimental systematic errors also include the uncertainties in the
determination of $P(c\rightarrow D^{*+})$ and $P(c\rightarrow D^0)$.

\begin{table}[t]\centering
\begin{tabular}{lrr}\hline\noalign{\smallskip}
Systematic Errors&&\cr
on $F_2^{c\overline c}$&$D^0$&$D^{*+}$\cr
\noalign{\smallskip}\hline\noalign{\smallskip}
fit &0.14&0.15\cr\noalign{\smallskip}
track efficiency&${+0.06\atop-0.02}$&${+0.09\atop-0.03}$\cr\noalign{\smallskip}
bin migrations&0.02&0.05\cr\noalign{\smallskip}
luminosity&0.015&0.015\cr
$\gamma p$ contribution&$<$0.004&$<$0.004\cr\noalign{\smallskip}
radiative corrections&0.03&0.03\cr\noalign{\smallskip}
fragmentation&&\cr
gluon splitting&0.02&0.02\cr\noalign{\smallskip}
$b$ contribution&$<$0.02&$<$0.02\cr\noalign{\smallskip}
charm fragmentation&&\cr
function&0.04&0.04\cr\noalign{\smallskip}
$P(c\rightarrow D)$&0.05&0.07\cr\noalign{\smallskip}
\hline\noalign{\smallskip}
total &${+0.17\atop-0.16}$
&${+0.20\atop-0.19}$\cr\noalign{\smallskip}\hline
\hline\noalign{\smallskip}
\end{tabular}
\caption{Summary of the average relative systematic uncertainties on the
measurement of $F_2^{c\overline c}$.}
\label{table3}
\end{table}
\begin{figure}[t]
\centering
\includegraphics[ bb= 0 140 566 410,width=16cm]{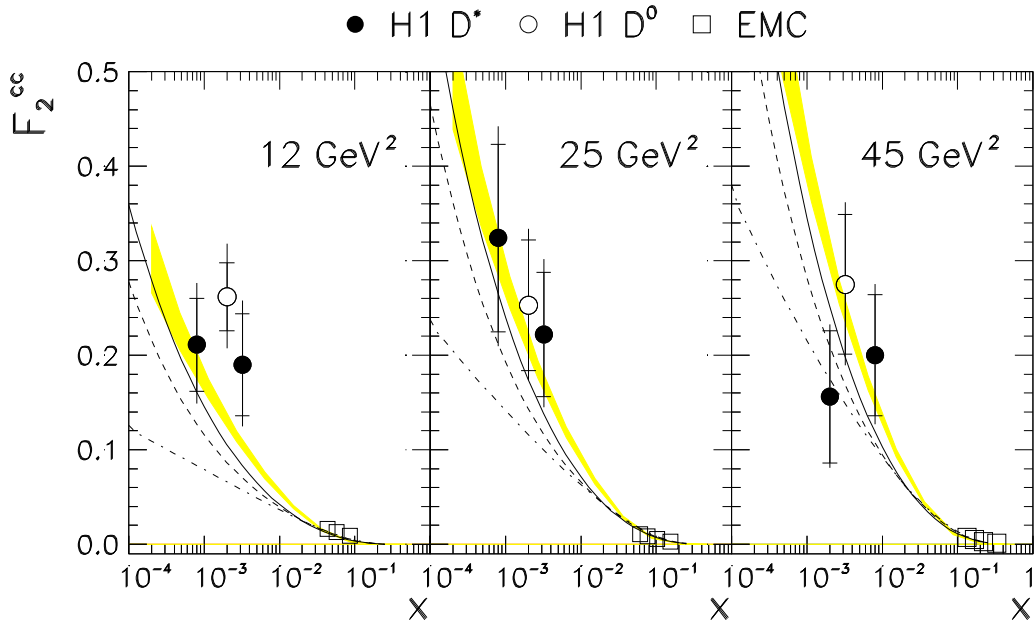}
\caption{ The charm contribution $F_2^{c\overline c}$ 
          to the proton structure function as derived from the inclusive
          $D^{*+}$ (full dots) and $D^0$ analysis (open circles)
 in comparison
          with the NLO calculations based on GRV-HO (full line), 
          MRSH (dashed line), and MRSD0$^\prime$ (dash-dotted line)
          parton distributions
          using a charm quark mass of $m_c=1.5\;\gev$ for 
          $\langle Q^2\rangle=12,\;25\;{\rm and }~45\;\gev^2$. 
          The inner (outer) error bars refer to the statistical 
          (total) errors. The shaded band represents the prediction from the 
          H1 NLO fit to the $F_2$ measurements.
          The EMC data are also shown (open boxes).
          }
\label{f2cplot}
\end{figure}

The measurement of $F_2^{c\overline c}$ is found to be fairly insensitive to 
the actual choice of the parameterization of the gluon density in the proton. 
Using GRV-HO or MRSD0$^\prime$ parameterization of the parton densities
 instead of MRSH,
changes the result by less than 2\% and 5\%, respectively.

Table~\ref{tf2c} also includes the ratio $F_2^{c\overline c}/F_2$ 
where the 
$F_2(x,Q^2)$ measurement is taken
from ref. \cite{h12}. Within the accuracy of the charm
data this ratio seems to be constant in the kinematic range explored in the
current analysis. 
In particular the steep rise of $F_2^{c\overline c}/F_2$ with
$Q^2$ for fixed $x$, as found by the EMC experiment \cite{emc} at 
considerably larger $x$, is not observed at HERA. The average yields
\begin{equation}
\displaystyle
\left\langle F_2^{c\overline c}/F_2\right\rangle=
0.237\pm0.021\;{+0.043\atop-0.039}\;, 
\end{equation}
which is an increase of about one order of magnitude
of the overall charm contribution compared to the EMC result. This is
consistent with the measured rise of the gluon distribution towards low $x$ 
and the dominance of the BGF process observed here.  
The present result on $F_2^{c\overline c}/F_2$ is also found to be in range
predicted by ref. \cite{kisselev}. 

The $F_2^{c\overline c}$ measurements are displayed in Fig. \ref{f2cplot}
together with the result of the EMC collaboration \cite{emc}.
The measurement at HERA extends the range of the
$F_2^{c\overline c}$ measurement
 by two orders of magnitude towards smaller $x$ values. 
The comparison of the H1 and EMC measurements reveals
a steep rise of $F_2^{c\overline c}$
with decreasing $x$. The data are
compared with  NLO calculations \cite{riemersma} using the GRV-HO, the 
MRSH, and the MRSD0$^\prime$ parameterizations of the gluon density in the 
proton for a charm quark mass
$m_c=1.5\;\gev$.
So far no experimental information is available on the gluon density in the proton
observed in charm production at small $x$. 
Therefore, also the MRSD0$^\prime$ parameterization is considered here, 
although it has already been excluded by inclusive measurements at HERA
\cite{h1first}.
This parameterization also fails to describe $F_2^{c\overline c}$.  
The data lie systematically above all predictions 
which is to be expected from the measurement of the charm production cross section. 

The data are also compared to the prediction from the H1 QCD fit to the 
$F_2$ measurements using a charm quark mass of $m_c=1.5\;\gev$. The error
band shown for this fit includes the propagation of the statistical and the
uncorrelated systematic errors on the total $F_2$ data through the fitting
procedure. This 
prediction is systematically above all other calculations, independently of $x$ 
and $Q^2$ but agrees better with the $F_2^{c\overline c}$ measurement.
The charm data indicate a small excess of
the measured $F_2^{c\overline c}$ with respect to all predictions
in the lowest $Q^2$ bin.

The dominant uncertainty in the QCD calculations arises from the uncertainty
in the charm quark mass. This affects mainly the lowest $Q^2$ bin, for which
a variation of $m_c$ by 200 MeV will change the prediction  
for $F_2^{c\overline c}$ by 15\%. However, the measurement of 
$F_2^{c\overline c}$ shows the same $m_c$ dependence. Therefore
the comparison of the data with QCD calculations based on different gluon 
densities in the proton is nearly independent on the assumption made for the 
charm quark mass.

\section{Conclusions}
Results on inclusive $D^0$ and $D^{*+}$ meson 
production in neutral current deep inelastic $ep$ scattering 
at HERA have been presented.
It has been shown, that the production dynamics of
charmed mesons in the current and central fragmentation region
 may  be described by the boson gluon fusion process.

The observed inclusive cross section ratio 
${\sigma(e p \rightarrow D^{*+} X)}/{\sigma(e p \rightarrow D^0 X)}$
is consistent with the results from $e^+e^-$ and hadroproduction data.

From the inclusive $D^0$ and $D^{*+}$ cross sections
a charm production cross section in deep inelastic $ep$ scattering for 
10~GeV$^2<Q^2<100$~GeV$^2$ and $0.01\le y\le0.7$ of
$\sigma\left(ep\rightarrow c\overline cX\right)=
(17.4\pm1.6\pm1.7\pm1.4)$~nb has been derived. The data have been compared to 
different NLO calculations including the H1 QCD fit to the total
$F_2$ data. The charm production cross section is found to be 
somewhat larger
than predicted.  

A first measurement of the charm contribution
$F_2^{c\overline c}\left(x,Q^2\right)$ to the
proton structure function for Bjorken $x$ between $8\cdot10^{-4}$ and
$8\cdot10^{-3}$ has been performed. 
Comparison of the present result with the EMC data reveals a
steep rise of $F_2^{c\overline c}\left(x,Q^2\right)$ with decreasing $x$. 
Agreement is observed between the measured $F_2^{c\overline c}$ and the
result of the NLO QCD fit of H1 to the inclusive $F_2$ data. 
Averaged over the kinematic range a ratio
$\left\langle F_2^{c\overline c}/F_2\right\rangle=
0.237\pm0.021\;{+0.043\atop-0.039}$ is obtained, which is one order of 
magnitude larger than at larger $x$.


\section*{Acknowledgments}

We gratefully acknowledge the efforts of the HERA machine group, 
the DESY computer center,
and the DESY directorate for their support.
The non-DESY members of the collaboration
wish to thank the DESY directorate for the
hospitality extended to them while working at DESY.
We would like to thank T. Behnke and M. Elsing for helpful discussions on the
heavy flavor results from LEP and A. Vogt for fruitful 
discussions on charm production at HERA.

\end{document}